\documentclass[a4paper,11pt]{article}
\pdfoutput=1
\usepackage{graphicx}
\usepackage{epsf,amsmath,bbold,amsfonts,stmaryrd}

\usepackage[utf8]{inputenc}
\usepackage{mathrsfs}
\usepackage{appendix}
\usepackage{amssymb}
\usepackage{float}
\usepackage{color}
\usepackage{cite}
\usepackage{hyperref}
\hypersetup{pageanchor=false}
\usepackage{indentfirst}
\usepackage{url}
\usepackage{float}
\usepackage{caption}
\usepackage[numbers,square,comma,sort&compress,merge]{natbib}

\newcommand{\be}{\begin{equation}}
\newcommand{\ee}{\end{equation}}
\newcommand{\bea}{\setlength\arraycolsep{2pt} \begin{eqnarray}}
\newcommand{\eea}{\end{eqnarray}}

\def\ft#1#2{{\textstyle{\frac{\scriptstyle #1}{\scriptstyle #2} } }}

\def\0{{\sst{(0)}}}
\def\1{{\sst{(1)}}}
\def\2{{\sst{(2)}}}
\def\3{{\sst{(3)}}}
\def\4{{\sst{(4)}}}
\def\5{{\sst{(5)}}}
\def\6{{\sst{(6)}}}
\def\7{{\sst{(7)}}}
\def\8{{\sst{(8)}}}
\def\sst#1{{\scriptscriptstyle #1}}

\begin{document}
\title{\textbf{Observational Signature and Additional Photon Rings of Asymmetric Thin-shell Wormhole}}
\author{Jun Peng$^{1}$,
Minyong Guo$^{2}$ and Xing-Hui Feng$^{3*}$}
\date{}
\maketitle

\vspace{-10mm}

\begin{center}
{\it$^1$Van Swinderen Institute, University of Groningen, 9747 AG Groningen, The Netherlands\\\vspace{1mm}

$^2$Center for High Energy Physics, Peking University,
 Beijing 100871, China\\\vspace{1mm}

$^3$Center for Joint Quantum Studies and Department of Physics,
School of Science, Tianjin University, Tianjin 300350, China\\\vspace{1mm}
}
\end{center}

\vspace{8mm}

\begin{abstract}
Recently, a distinct shadow mechanism was proposed by Wang et al. from the asymmetric thin-shell wormhole (ATW) in [Phys. Lett. B 811 (2020) 135930]. On the other hand, Gralla et al's work [Phys. Rev. D 100 (2019) 2, 024018] represented a nice description of photon rings in the presence of an accretion disk around a black hole. In this paper, we are inspired to thoroughly investigate the observational appearance of accretion disk around the ATW. Although the spacetime outside an ATW with a throat could be identical to that containing a black hole with its event horizon, we show evident additional photon rings from the ATW spacetime. Moreover, a potential lensing band between two highly demagnified photon rings is found. Our analysis provides an optically observational signature to distinguish ATWs from black holes.
\end{abstract}

\vfill{\footnotesize jun.peng@rug.nl,\,\, minyongguo@pku.edu.cn,\,\, xhfeng@tju.edu.cn,\\$~~~~~~*$ Corresponding author.}

\maketitle

\newpage

\section{Introduction}
The first black hole image of M87* released by the Event Horizon Telescope plays a significant role in the frontier of general relativity \cite{Akiyama:2019cqa}. It provides a direct and powerful observational information for general relativity in the strong gravitational regime. As a result the research of black hole shadow becomes very popular thereafter \cite{Cardoso:2008bp, Zhang:2019glo, Cai:2021fpr, Bambi:2019tjh, Lu:2019zxb,Feng:2019zzn, Guo:2019lur,Ma:2019ybz,Yang:2019zcn,Guo:2020zmf,Zeng:2020dco,Zeng:2020vsj, Perlick:2018iye, Li:2020drn, Zhang:2020xub, Qian:2021qow, Wang:2017hjl, Wei:2013kza, Yang:2021zqy, Gralla:2019xty,Johnson:2019ljv,Himwich:2020msm,Gralla:2020yvo,Gralla:2020srx,Peng:2020wun}. Among these interesting works, Gralla et al. originality give an elegant description on  black hole shadows, lensing rings and photon rings considering an emission disk around black holes \cite{Gralla:2019xty}, which naturally led to lots of interesting follow-up works \cite{Johnson:2019ljv,Himwich:2020msm,Gralla:2020yvo,Gralla:2020srx,Peng:2020wun}. In this context, the term 'shadow' refers to a dark area outside the black hole. While, the regular `shadow' denotes the critical curve in the sky of observers and the critical curve is closely related to the spherical photon orbits which are always radially unstable \cite{Guo:2020qwk}. To avoid misunderstanding, for the latter we go by the name of critical curve in the rest of this article. Therefore, one can conclude that the shape of the shadow could change if different sources of light are considered for the same black hole. On the contrary, the critical curve is invariant as long as the spacetime geometry are given.

Nevertheless, with the help of the image of M87* we are still not enough to assert that the supermassive object in the center of M87* must be a black hole. This is because black holes are not the only ones that have photon spheres which have the ability to shade the horizon. It has been found some ultra compact objects (UCOs) also own photon spheres \cite{Guo:2020qwk, Cunha:2017qtt}. Furthermore, some of them are found to mimic the optical appearances of black holes including shadows \cite{Abdikamalov:2019ztb, Narzilloev:2020peq, Herdeiro:2021lwl}. Thus it is crucial to distinguish black holes from UCOs whose external spacetime geometries are similar or even identical to those of black holes up to the vicinity of horizon, such as gravastar, wormhole and fuzzball and so on. In fact one may distinguish black holes from some UCOs in their acoustic properties, such as echo effect \cite{Konoplya:2018yrp,Cardoso:2019rvt,Maggio:2020jml,Buoninfante:2019teo}. Roughly speaking, the essential distinction between black holes and UCOs is that the event horizon of black holes is a one-way membrane while UCOs have no horizons. In this sense, black holes can be regarded as absolute black objects with zero reflection, while UCOs have considerable reflection.

Recently the shadow of an asymmetric thin-shell wormhole (ATW) has been studied in \cite{Wang:2020emr}. The authors showed that due to reflection of photons by the wormhole there exist a novel shadow which is different from that of black hole in certain parameters space. Then more examples of asymmetric shin-shell wormhole are studied in \cite{Guerrero:2021pxt, Tsukamoto:2021fpp, Wielgus:2020uqz}, of which the authors named their results as double shadows \cite{Wielgus:2020uqz}. It's worth noting that double shadows should be called double critical curves or photon spheres in the context of our article. On the other hand, the novel shadow proposed in \cite{Wang:2020emr} does refer to the dark area, however, its edge is still related to the photon sphere when the throat of the ATW is inside the photon sphere, but the photon sphere here is the one which is located in the other side opposite the observers other than the usual one in the observers' side for an ATW spacetime. So far there's no discussion on the observational appearance of emission disk in ATW spacetime although no matter novel shadows or double shadows are improper terminologies in practical observation. Our motivation is to give a completed picture of observational appearance of accretion disk around an ATW.

The plan of our paper is organized as follows: we first give a brief review of the asymmetric thin-shell wormhole in section II. Then we analysis the photon trajectories and deflection angles in this wormhole in section III. Next we discuss the transfer functions and observational appearances of emission disks around this wormhole in section IV. We conclude in section V.

\section{Null geodesic in asymmetric thin-shell wormhole}
In this section, we give a brief review of the asymmetric thin-shell wormhole model presented in \cite{Wang:2020emr}. We consider a thin-shell wormhole using cut-and-paste method, that is, two distinct spacetimes with different parameters are glued by a thin shell
\be
ds_i^2=-f_i(r_i)dt_i^2+\frac{dr_i^2}{f_i(r_i)}+r_i^2d\Omega^2,
\ee
where $i=1,2$, and by focusing on the Schwarzschild case we have
\be
f_i(r_i)=1-\frac{2M_i}{r_i},\quad r\ge R,
\ee
where $M_i$ are the mass parameters, and $R$ is the position of thin-shell, i.e. the radius of throat. Thus we have
\be
R>{\rm max}\{2M_1,2M_2\},
\ee

\begin{figure}[h!]
\begin{center}
\includegraphics[scale=0.5]{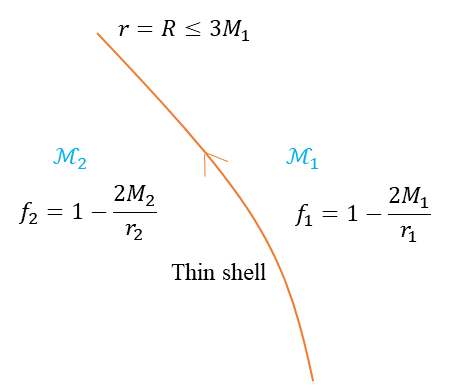}
\caption{Asymmetric shin-shell wormhole as a black hole mimicer. We assume that observers are located at spacetime ${\cal M}_1$. So the spacetime geometry viewed by observers is identical to that of black hole up to the thin-shell.}\label{potential}
\end{center}
\end{figure}
In each spacetime, the radial null geodesic is
\be
\Big(\frac{dr_i}{d\tau}\Big)^2+V_{i,eff}=\frac{1}{b^2_i}
\ee
with the effective potential given by
\be
V_{i,eff}=\frac{f_i(r_i)}{r_i^2}
\ee
where $b_i=\ft {L_i}{E_i}$ is called the impact parameter.

Without loss of generality, we suppose that observers are located in spacetime ${\cal M}_1$ and set $M_1=1$ and $M_2=k$. The impact parameters in two spacetimes are connected with each other by
\be
\frac{b_1}{b_2}=\sqrt{\frac{R-2k}{R-2}}\equiv Z\label{relation}
\ee
If there exists photon sphere in ${\cal M}_1$, it is hardly to distinguish a wormhole from a black hole based on direct optical observation, because the horizon is shaded by the photon sphere. So we are interested in this case, i.e. $R<3$. In other words, we want to distinguish a wormhole from a black hole through more abundant information even if they look like each other.

The more abundant information indeed exists. Considering the ingoing null geodesics with $b_1<b_1^c=3\sqrt3$ in spacetime ${\cal M}_1$, certain geodesics would turn back passing through the throat with a necessary condition
\be
b_2=\frac{b_1}{Z}>b_2^c=3\sqrt3k.
\ee
To summarize, the ingoing null geodesic in spacetime ${\cal M}_1$ whose impact parameter satisfies
\be
3\sqrt3kZ<b_1<3\sqrt3,
\ee
would drop into spacetime ${\cal M}_2$ and then turn back to spacetime ${\cal M}_1$ passing through the throat. This condition can be satisfied given
\be
1<k<\frac R2\le\frac32.
\ee

\begin{figure}[t!]
\begin{center}
\includegraphics[scale=0.5]{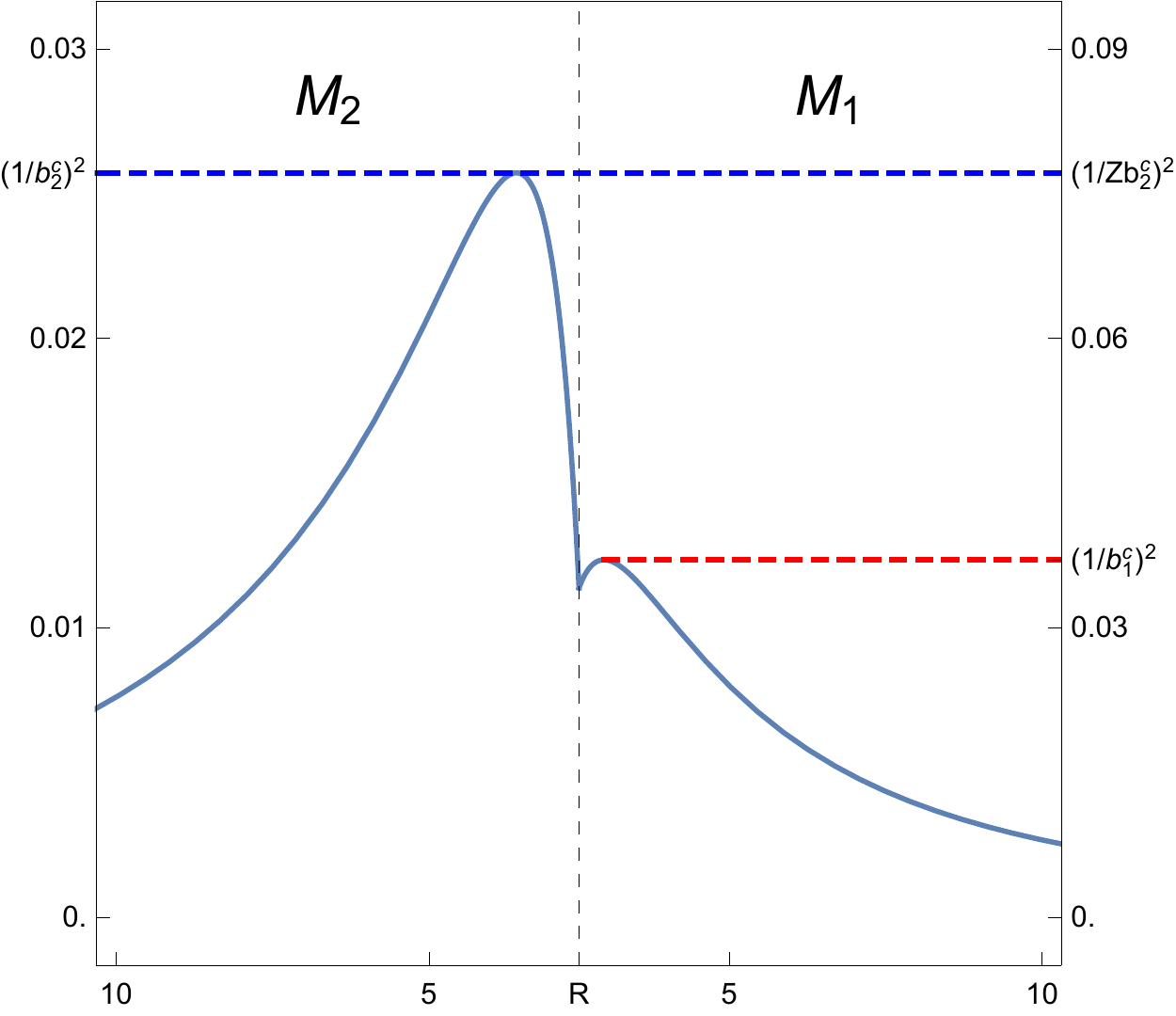}
\caption{Plot of effective potential of radial null geodesic in asymmetric thin-shell wormhole. We have set $M_1=1, M_2=1.2, R=2.6$, so $b_1^c=3\sqrt3\approx5.19615$ and $Zb_2^c=3.6$ here and in the following. Note that the ticks in spacetime ${\cal M}_2$ is scaled by $Z^2$ because the impact parameters in two spacetimes are connected with each other by \eqref{relation}}\label{potential}
\end{center}
\end{figure}

We show a plot of effective potential in Fig.\ref{potential}. It can be obviously seen from this plot that the photon in spacetime ${\cal M}_1$ with impact parameter lying in the range $Zb_2^c<b_1<b_1^c$ can turn back when it reaches the turning point in spacetime ${\cal M}_2$. We will see that this reflection mechanism by wormhole make a essentially distinction in the observational appearance between wormhole and black hole in the following.

\section{Trajectory of photon and deflection angle in asymmetric thin-shell wormhole}
To have a complete understanding of the observational appearance of accretion disk around asymmetric thin-shell wormhole through the shadow, photon rings and lensing rings, we need first to investigate the trajectory and deflection angle of a light ray traveling in the wormhole. It is convenient to make a coordinate transformation $u_i=1/r_i$. The trajectory of photon is determined by orbit equation
\be
\Big(\frac{du_i}{d\phi}\Big)^2=G_i(u_i)\label{geodesic},
\ee
where
\be
G_i(u_i)=\frac{1}{b_i^2}+2M_iu_i^3-u_i^2.
\ee
\begin{figure}[t!]
\begin{center}
\includegraphics[scale=0.3]{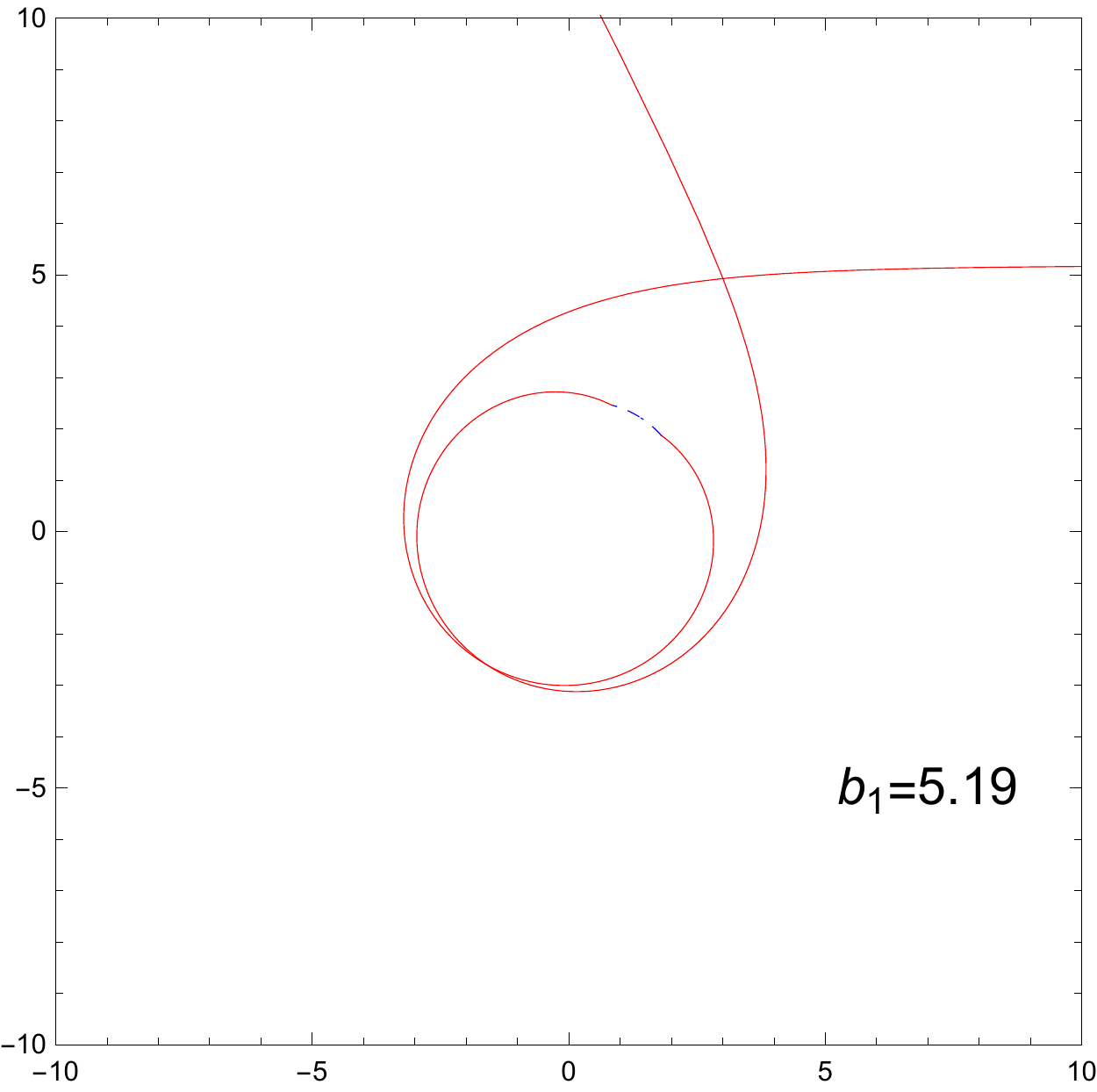}\qquad\includegraphics[scale=0.3]{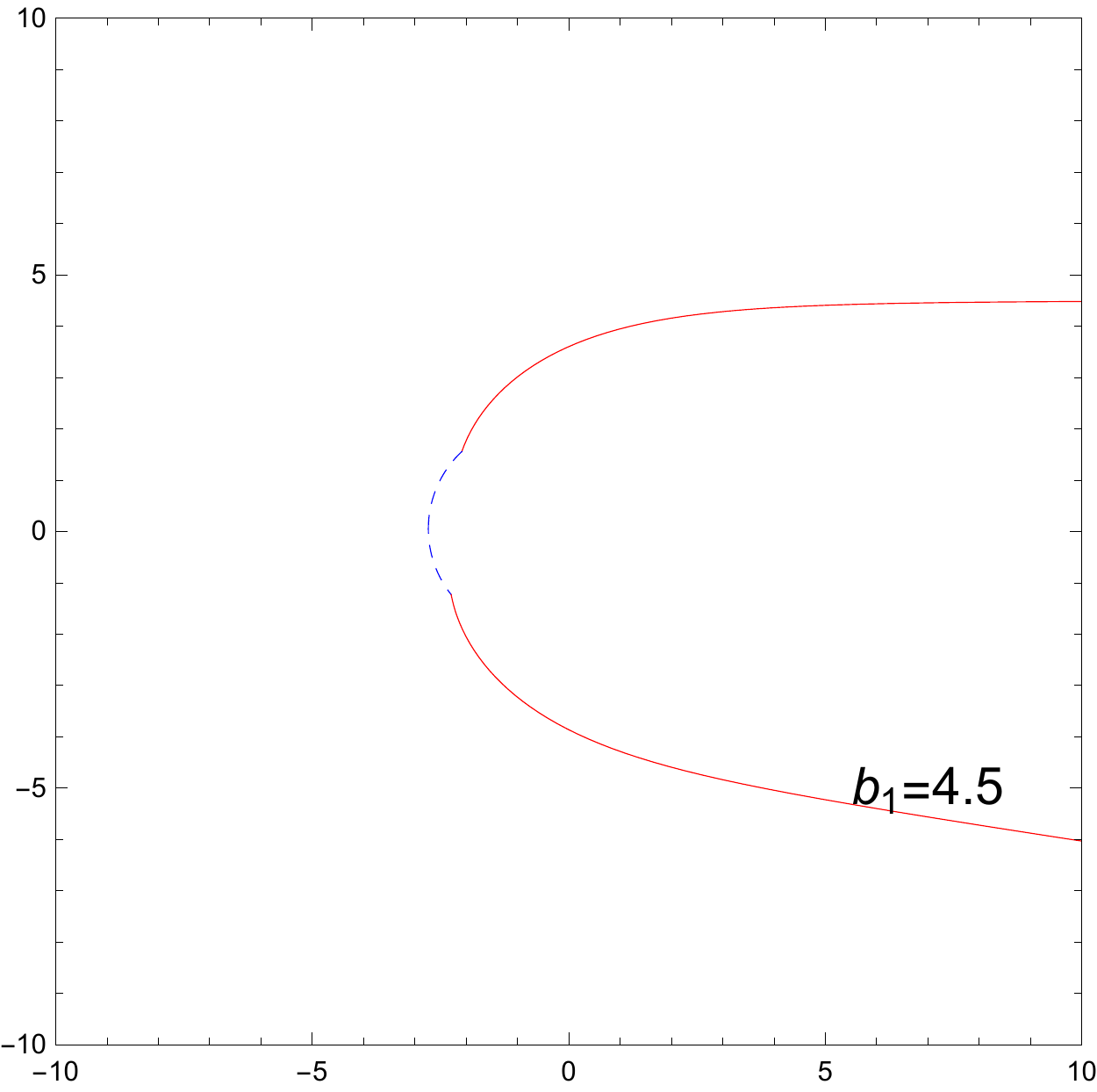}\qquad\includegraphics[scale=0.3]{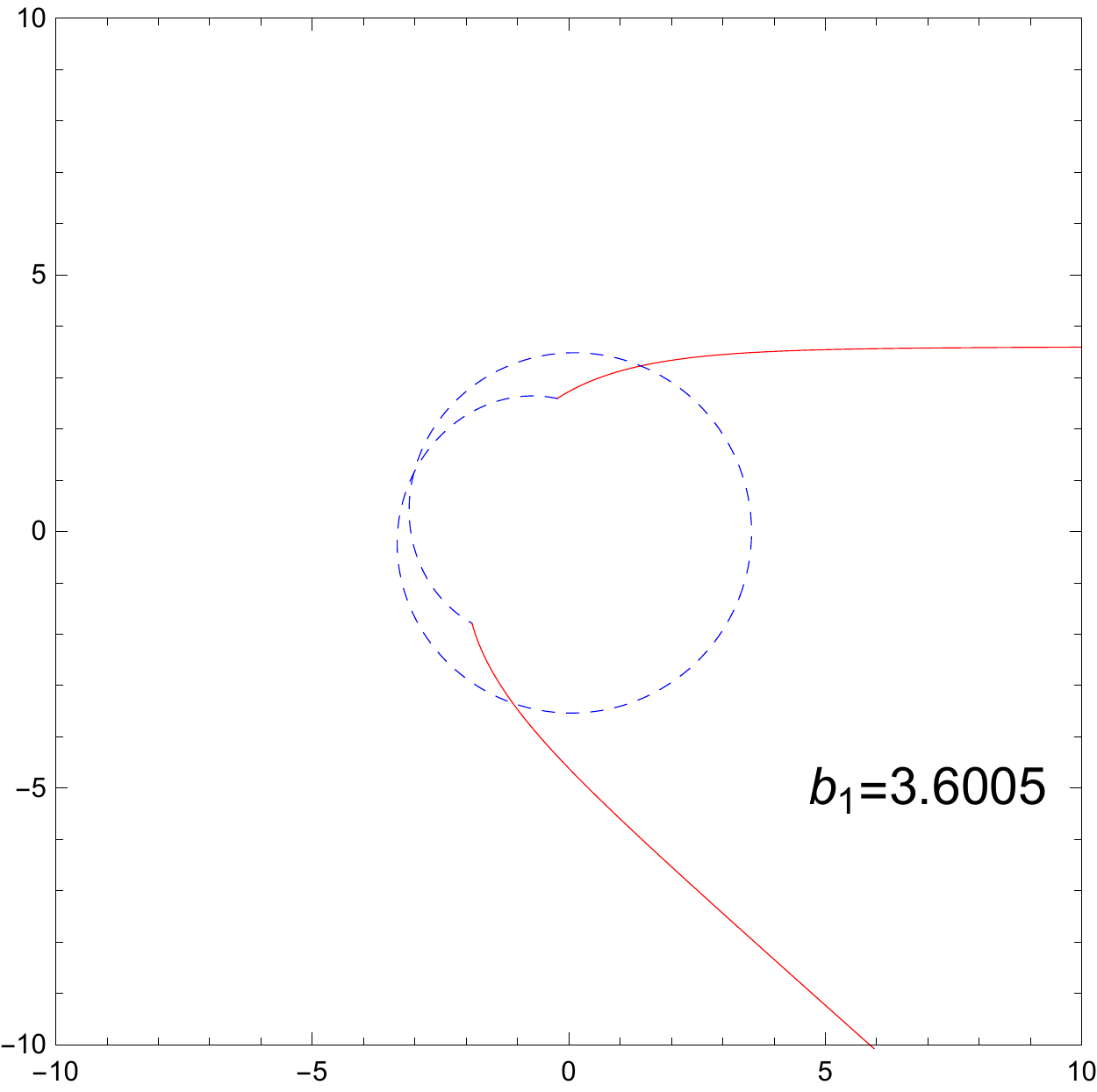}
\caption{Photon trajectories in the Euclidean polar coordinates $(r,\phi)$ with impact parameters lying in the range $Zb_2^c<b_1<b_1^c$. The photons are coming from far right in spacetime ${\cal M}_1$. The red solid lines stand for trajectories in spacetime ${\cal M}_1$, and the blue dashed lines stand for trajectories in spacetime ${\cal M}_2$.}.\label{trajectory}
\end{center}
\end{figure}
All trajectories can be divided into three classes. When $b_1>b_1^c$, photon in ${\cal M}_1$ from infinity approach one closest point outside the throat, and then move back to infinity in ${\cal M}_1$. When $Zb_2^c<b_1<b_1^c$, photon in ${\cal M}_1$ from infinity drop into ${\cal M}_2$ and then turn back passing through the throat to infinity in ${\cal M}_1$. When $b_1<Zb_2^c$, photon in ${\cal M}_1$ from infinity drop into ${\cal M}_2$ and move to infinity in ${\cal M}_2$.

For $b_1>b_1^c$, the turning point in spacetime ${\cal M}_1$ corresponds to the minimally positive real root of $G_1(u_1)=0$, which we will denote by $u_1^{min}$. According to \eqref{geodesic}, the total change of azimuthal angle $\phi$ for certain trajectory with impact parameter $b_1$ can be calculated by
\be
\phi_1(b_1)=2\int_0^{u_1^{min}}\frac{du_1}{\sqrt{G_1(u_1)}},\quad b_1>b_1^c.
\ee
For
\be
Zb_2^c<b_1<b_1^c,
\ee
we firstly focus on the trajectory in spacetime ${\cal M}_1$ outside the throat. The total change of azimuthal angle $\phi$ in spacetime ${\cal M}_1$ is obtained by
\be
\phi_1(b_1)=\int_0^{1/R}\frac{du_1}{\sqrt{G_1(u_1)}},\quad b_1<b_1^c,
\ee
The turning point in spacetime ${\cal M}_2$ corresponds to the maximally positive real root of $G_2(u_2)=0$, which we will denote by $u_2^{max}$. According to \eqref{geodesic}, the total change of azimuthal angle $\phi$ for certain trajectory with impact parameter $b_2$ in spacetime ${\cal M}_2$ can be calculated by
\be
\phi_2(b_2)=2\int_{u_2^{max}}^{1/R}\frac{du_2}{\sqrt{G_2(u_2)}},\quad b_2>b_2^c
\ee
It is convenient to define three orbit numbers
\be
n_1(b_1) = \frac{\phi_1(b_1)}{2\pi},\quad n_2(b_1) = \frac{\phi_1(b_1)+\phi_2(b_1/Z)}{2\pi},\quad n_3(b_1) = \frac{2\phi_1(b_1)+\phi_2(b_1/Z)}{2\pi}
\ee
for later use.

\section{Transfer functions and photon rings in asymmetric thin-shell wormhole}
As Gralla et al's proposal \cite{Gralla:2019xty}, we focus on the optically and geometrically thin accretion disks for simplicity. Both static observer and accretion disk are assumed to locate at spacetime ${\cal M}_1$. The static observer is assumed to locate at the north pole, and the accretion disk is on the equatorial plane. The lights emitted from the accretion disk is considered isotropic in the rest frame of the static observer. In view of the spherical symmetry of the spacetime, we also suppose the emitted specific intensity only depends on the radial coordinate, denoted by $I^{\rm em}_\nu(r)$ with emission frequency $\nu$ in a static frame. An observer in infinity will receive the specific intensity $I^{\rm obs}_{\nu'}$ with redshifted frequency $\nu'=\sqrt{f}\nu$. Considering $I_\nu/\nu^3$ is conserved along a ray, i.e.
\be
\frac{I^{\rm obs}_{\nu'}}{\nu'^3}=\frac{I^{\rm em}_\nu}{\nu^3}
\ee
we have the observed specific intensity
\be
I^{\rm obs}_{\nu'}=f^{3/2}(r)I^{\rm em}_\nu(r)
\ee
So the total observed intensity is an integral over all frequencies
\be
I^{\rm obs}=\int I^{\rm obs}_{\nu'}d\nu'=\int f^2 I^{\rm em}_\nu d\nu=f^2(r)I^{\rm em}(r)
\ee
where $I^{\rm em}=\int I^{\rm em}_\nu d\nu$ is the total emitted intensity from the accretion disk.

\begin{figure}[t!]
\begin{center}
\includegraphics[scale=0.3]{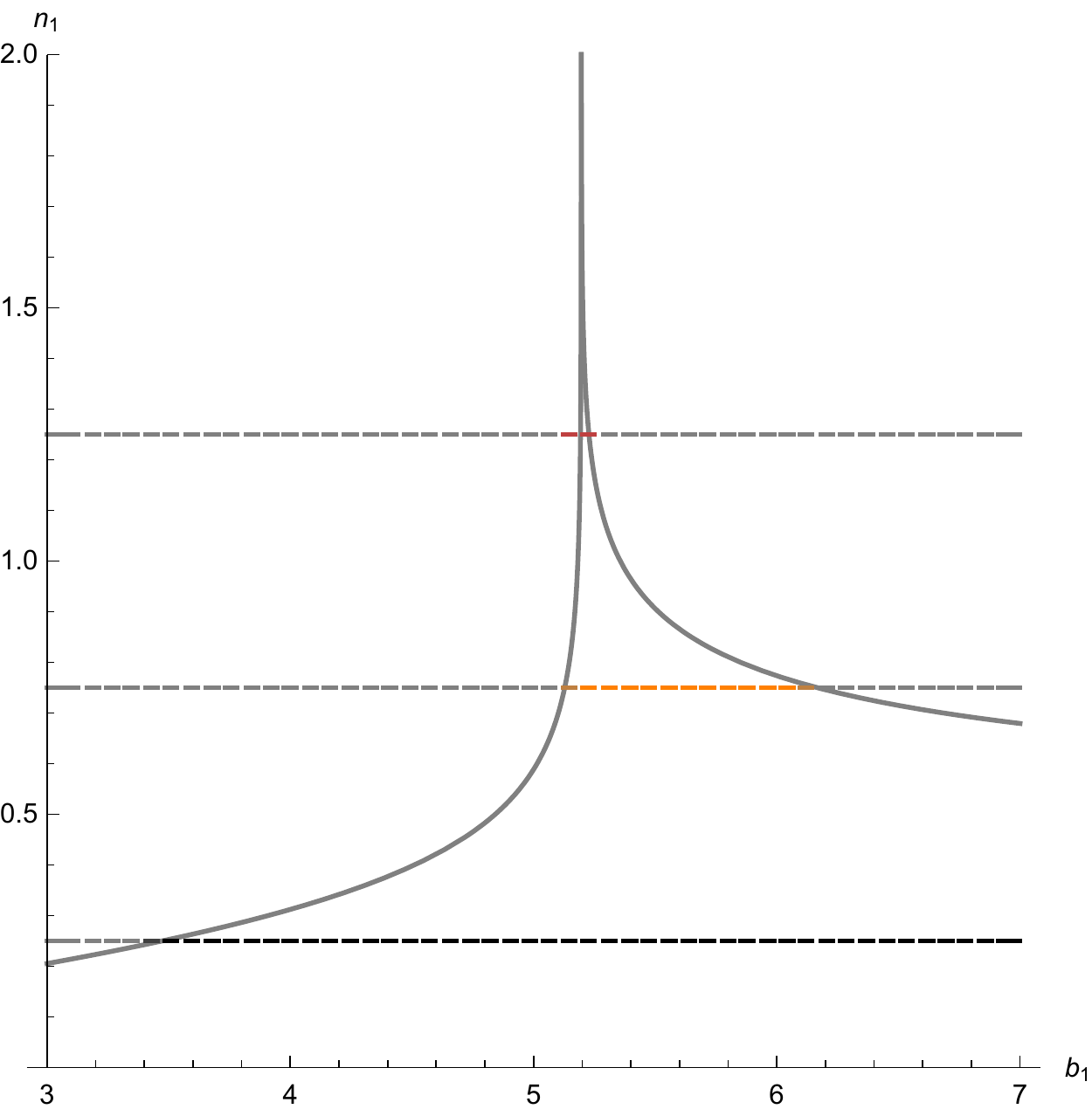}\qquad\includegraphics[scale=0.3]{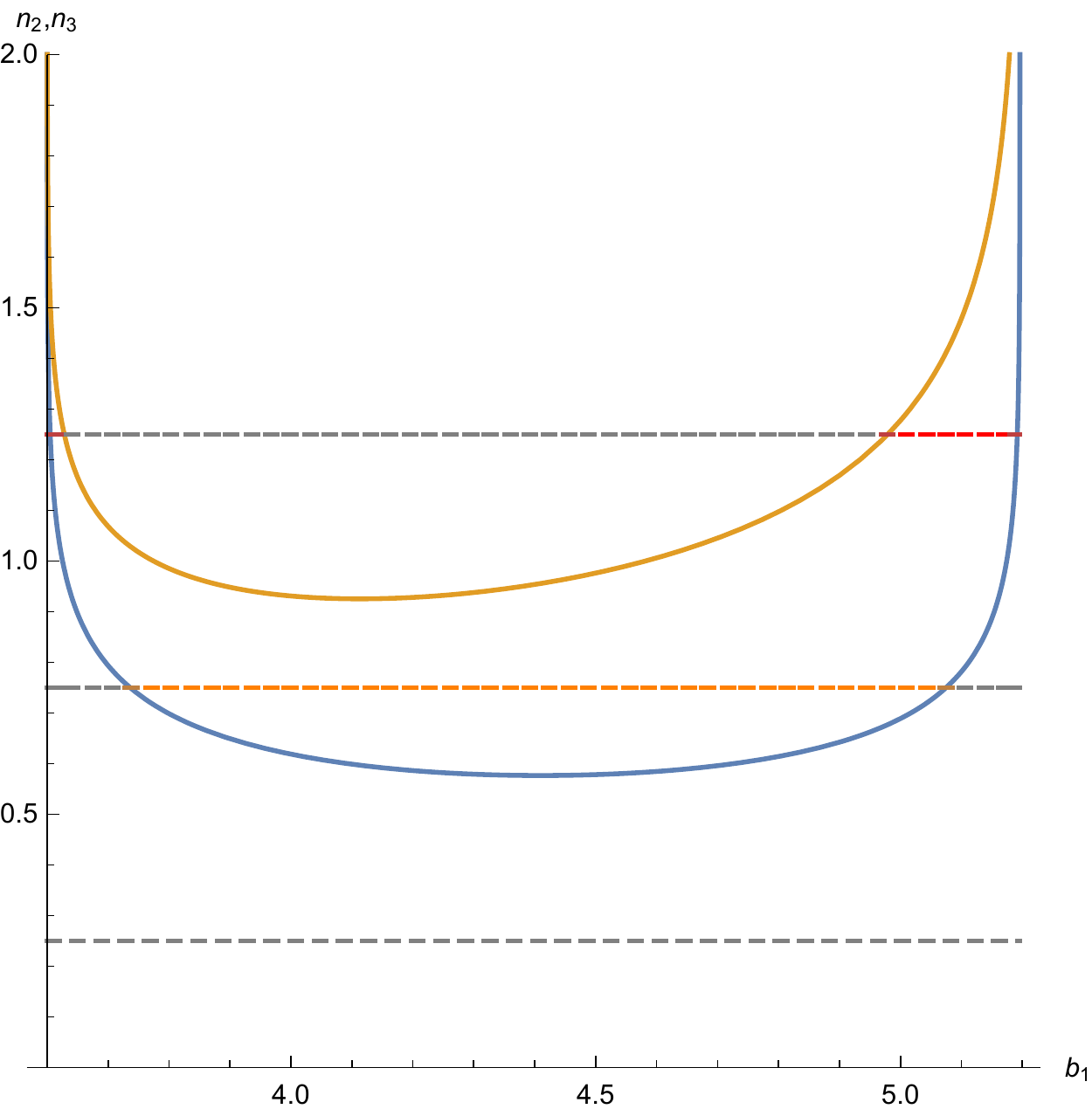}\qquad\includegraphics[scale=0.5]{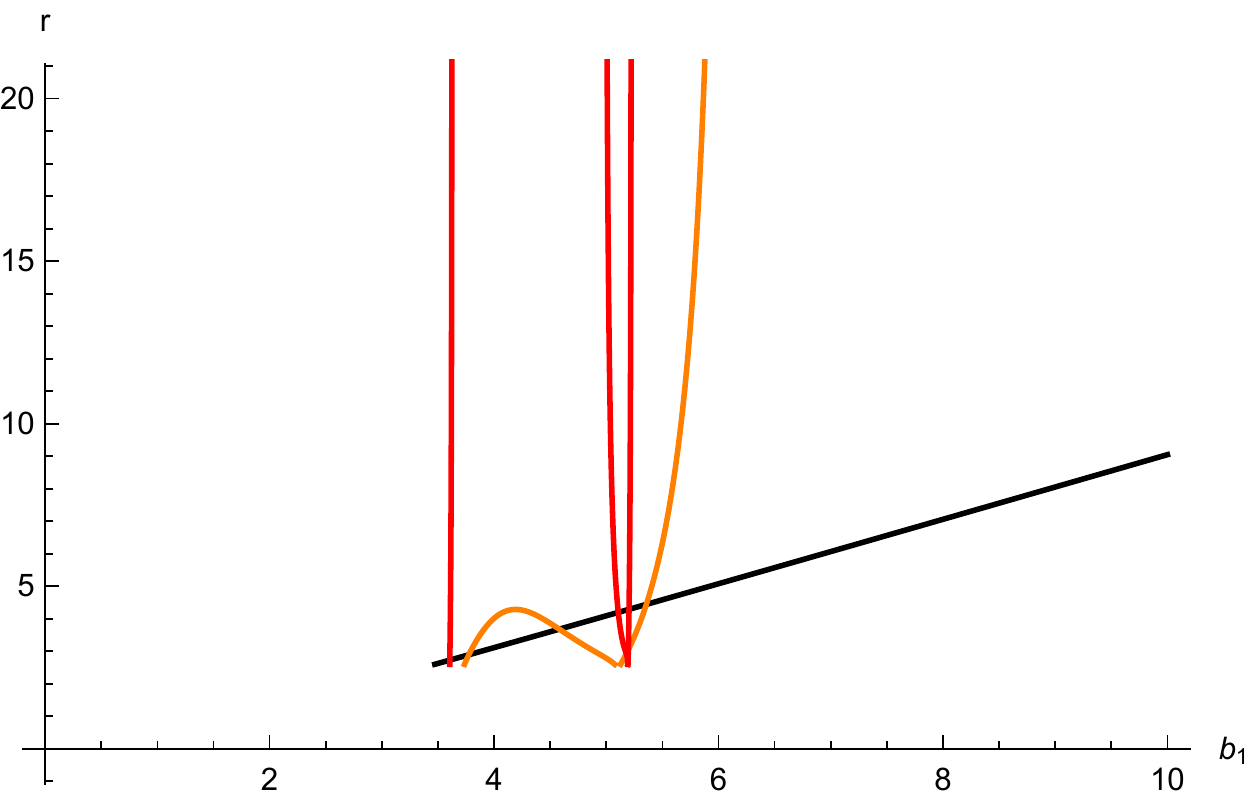}
\caption{The ranges of impact parameter and plots for first three transfer functions. The black, orange and red ones correspond respectively to the first, second and third transfer functions.}\label{transfer}
\end{center}
\end{figure}

According to ray-tracing method, if a light ray from the observer intersects with the emission disk, it means the intersecting point as a light source will contribute brightness to the observer. We first consider that a light ray completely travels in spacetime ${\cal M}_1$. As the black hole case, a light ray whose orbit number $n_1>1/4$ will intersect with the disk on the front side. If $n_1$ goes larger than $3/4$, the light ray will bend around the wormhole, intersecting with the disk for the second time on the back side. Further, when $n_1>5/4$, the light ray will intersect with the disk for the third time on the front side again, and so on. Hence, the observed intensity is a sum of the intensities from each intersection,
\be
I^{\rm obs}(b_1)=\sum_m f^2I^{\rm em}|_{r=r_m(b_1)}\label{Iobs}
\ee
where $r_m(b_1)$ is the so called transfer function which denotes the radial position of the $m$-th intersection with the emission disk.

As discussed in above sections, for asymmetric thin shell wormhole photon can turn back passing through the throat in proper parameters space, so we would have addition transfer functions. According to the definitions of orbit numbers, when $n_2<3/4$ and $n_3>3/4$, the reflected outgoing trajectories in spacetime ${\cal M}_1$ will intersect with the disk on the back side. When $n_2<5/4$ and $n_3>5/4$, the reflected outgoing trajectories in spacetime ${\cal M}_1$ will intersect with the disk on the front side. Thus the impact parameter range for new second transfer functions are determined by $n_2<3/4$ and $n_3>3/4$ and the impact parameter range for new third transfer functions are determined by $n_2<5/4$ and $n_3>5/4$ as illustrated in the middle plot of Fig.\ref{transfer}. We give all transfer functions in the right plot of Fig.\ref{transfer}.

As illustrated in \cite{Gralla:2019xty}, the first transfer function gives the "direct image" of the disk which is essentially just the redshift of source profile. The second transfer function gives a highly demagnified image of the back side of the disk, referred to "lensing ring". The third transfer function gives an extremely demagnified image of the front side of the disk, referred to "photon ring". The images resulted from further transfer functions are so demagnified that they can be neglected. The demagnified scale is determined by the slope of transfer function, $dr/d\phi$, called the demagnification factor. We can see from the right plot of Fig.\ref{transfer} that the new third transfer function near $Zb_2^c$ has a high slope like the usual third transfer function near $b_1^c$. Another new third transfer function in the left of $b_1^c$ has a smaller slope than the usual third transfer function near $b_1^c$, but bigger slope than the usual second transfer function in the right of $b_1^c$. The new second transfer function has a modest slope like the usual first transfer function, so we had better call the resulted image as 'lensing band'.

Now we take two typical emission models as examples proposed by Gralla et al. \cite{Gralla:2019xty} to make physical pictures more clear

\begin{figure}[h!]
\begin{center}
\includegraphics[scale=0.5]{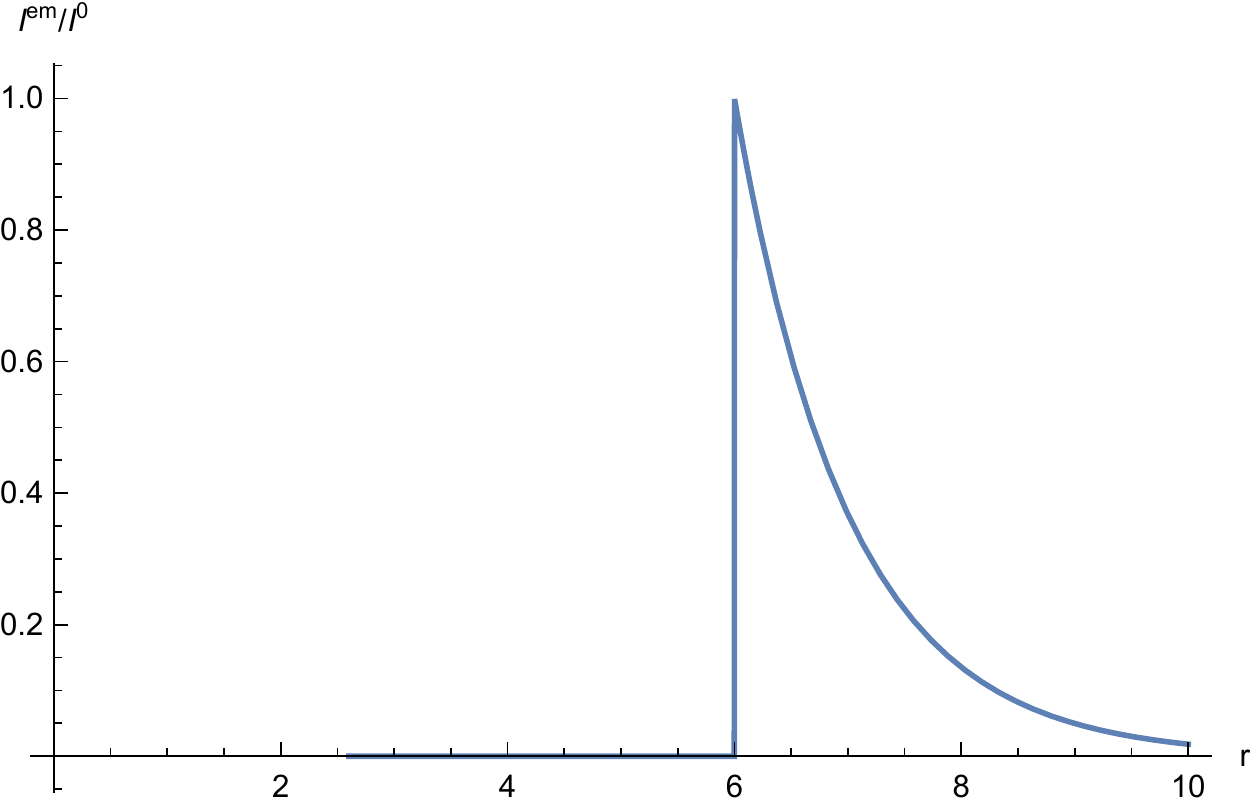}\qquad\includegraphics[scale=0.5]{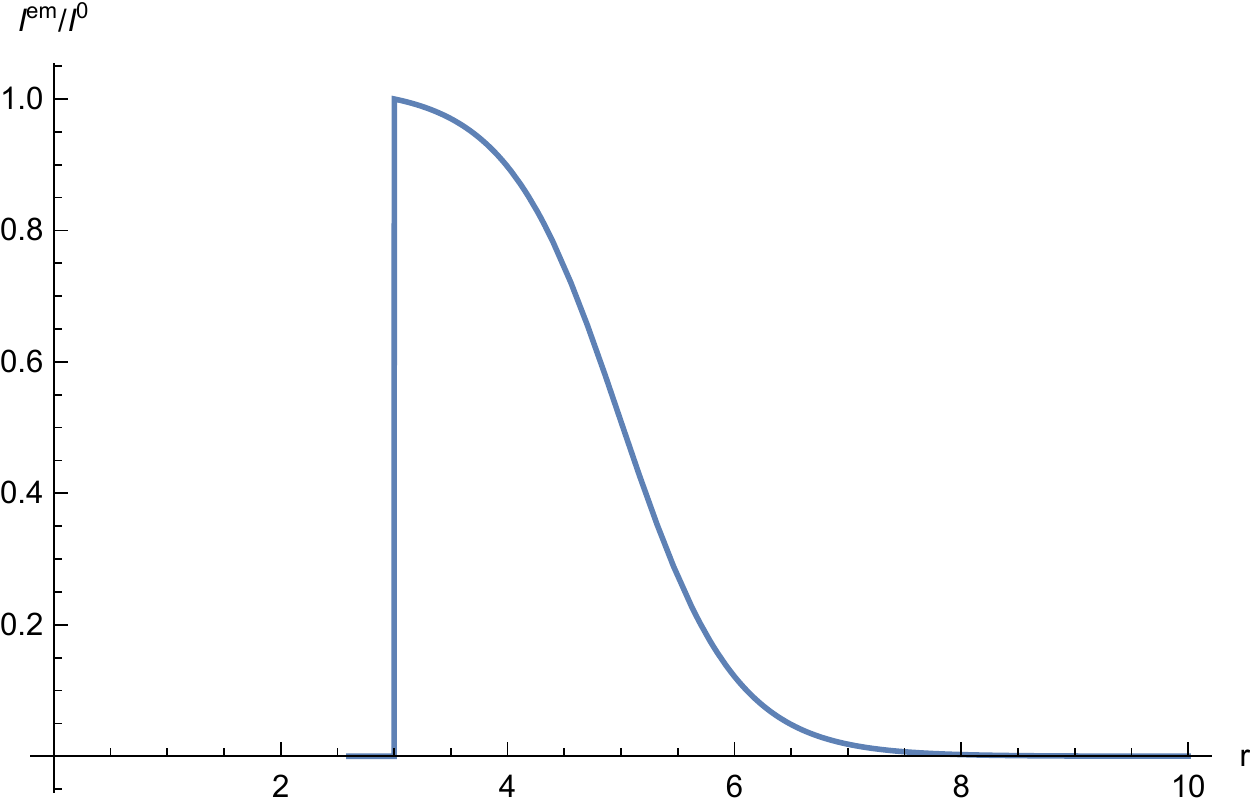}
\caption{The left emission profile (model I) is sharply peaked and abruptly end at the innermost stable circular orbit ($6M_1$). The right emission profile (model II) decaying gradually from the photon sphere ($3M_1$) to the innermost stable circular orbit ($6M_1$)}
\end{center}
\end{figure}

\begin{figure}[h!]
\begin{center}
\includegraphics[scale=0.45]{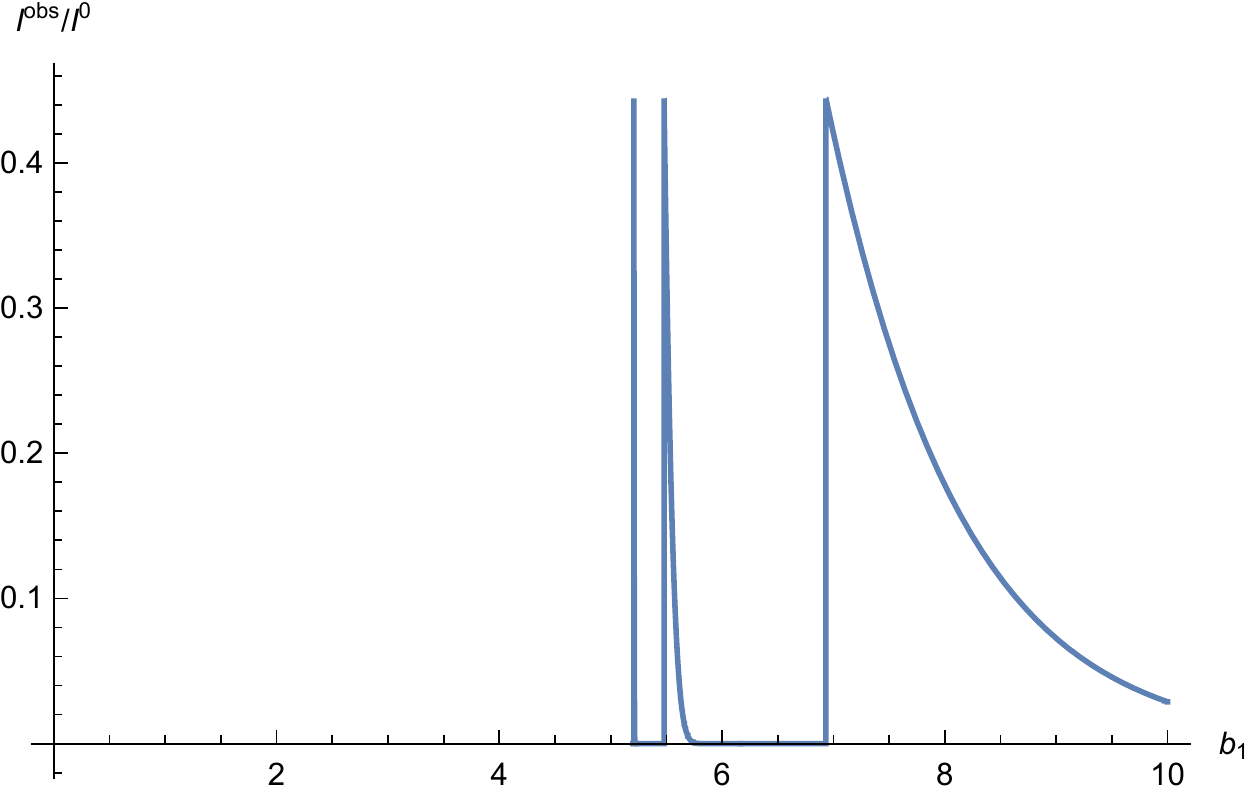}\qquad\includegraphics[scale=0.4]{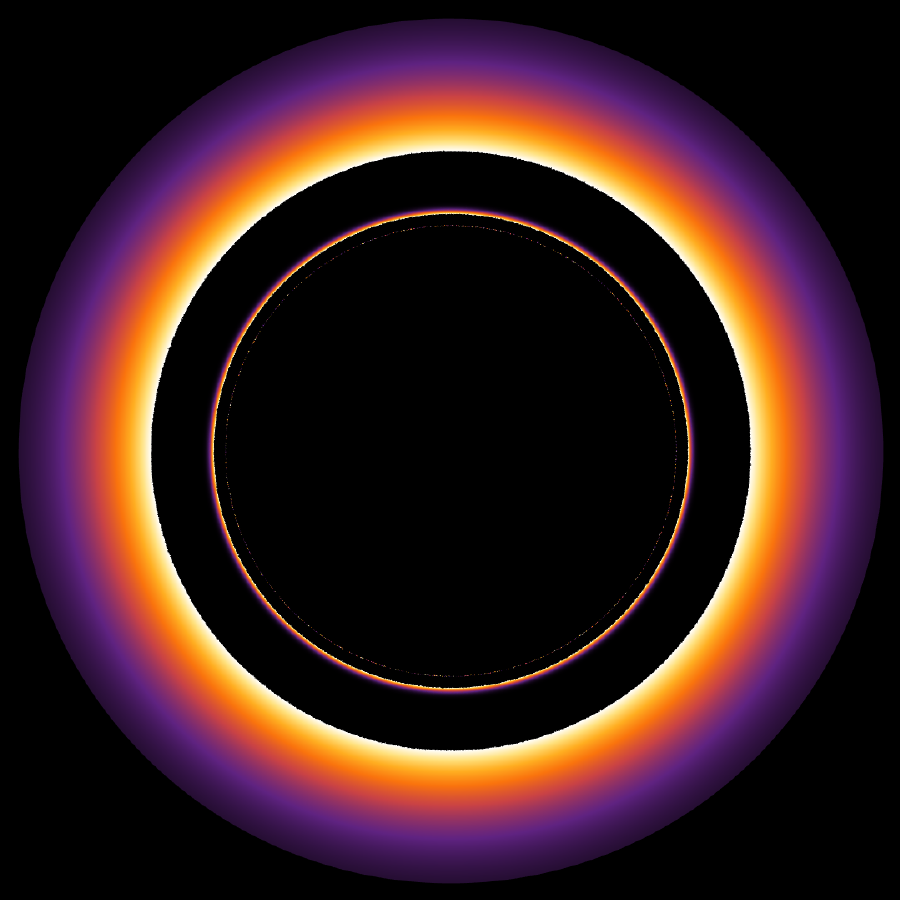}\qquad\includegraphics[scale=0.4]{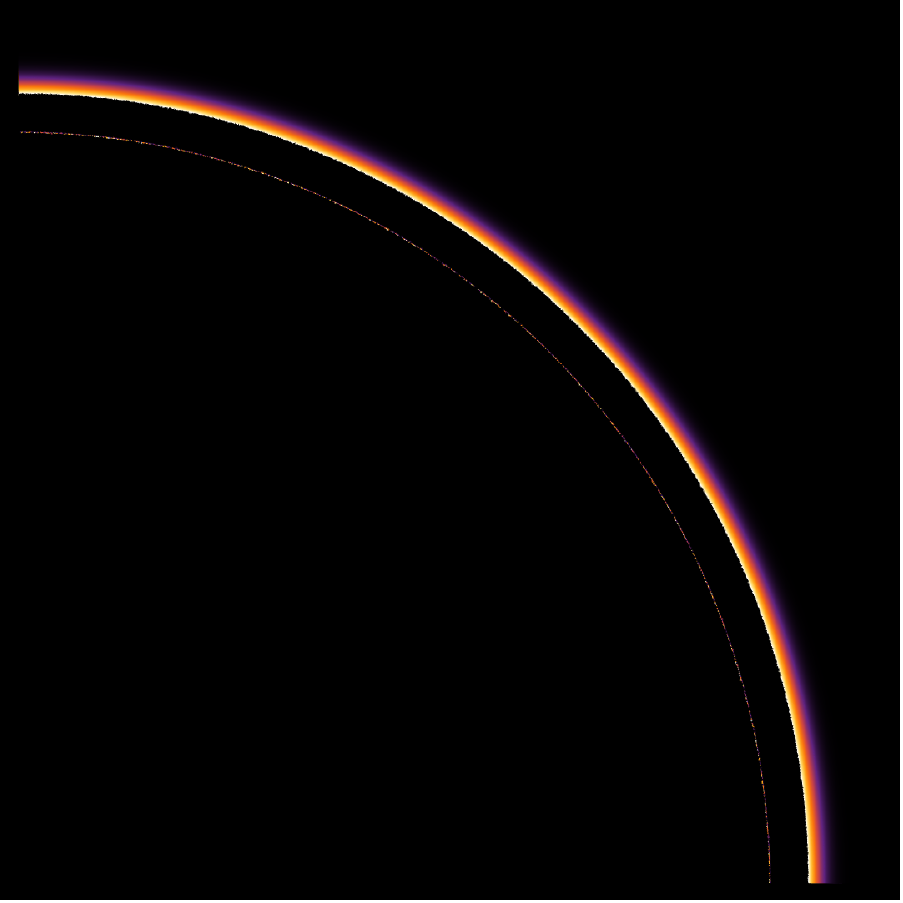}\\
\includegraphics[scale=0.45]{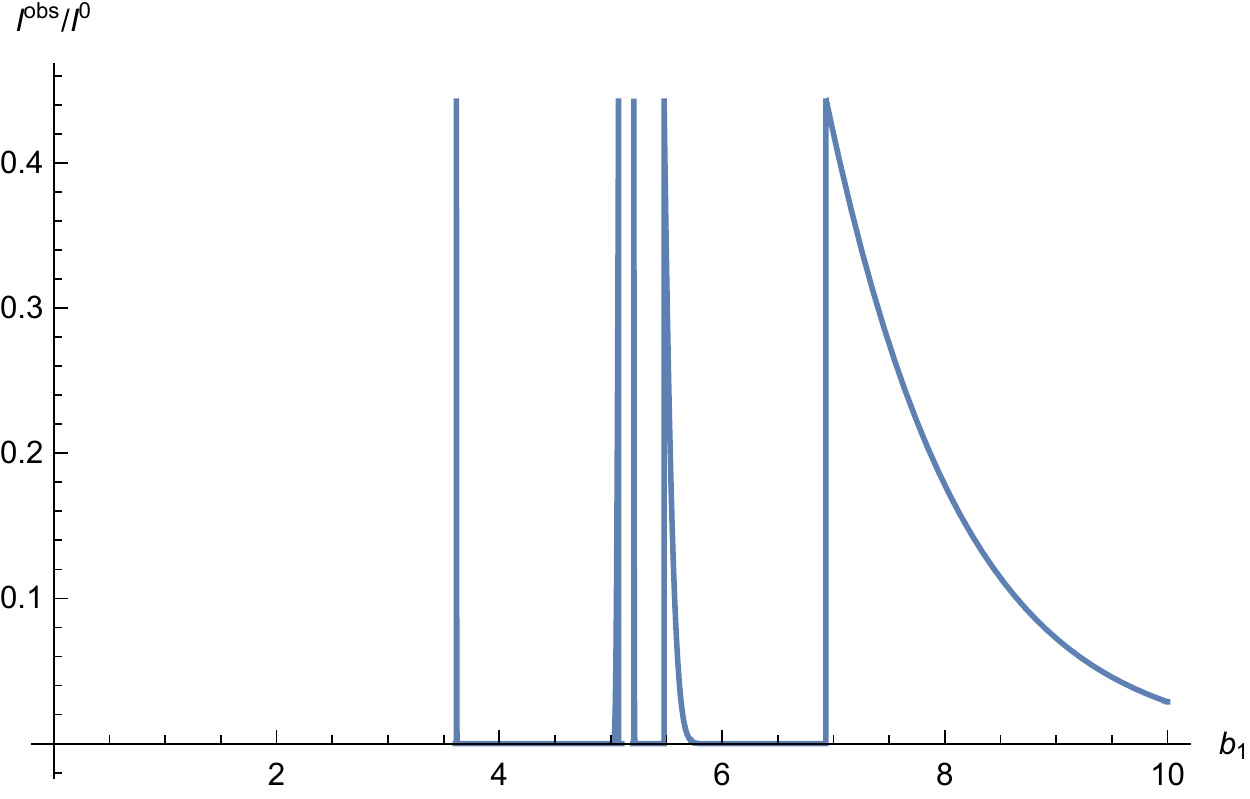}\qquad\includegraphics[scale=0.4]{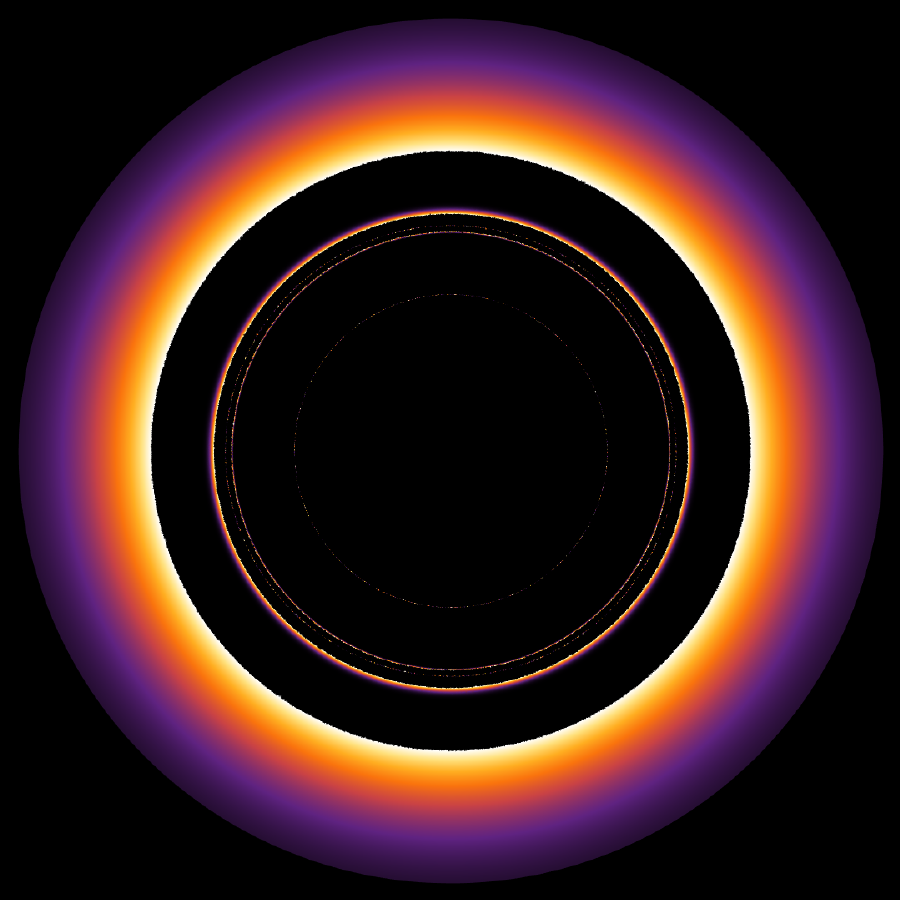}\qquad\includegraphics[scale=0.4]{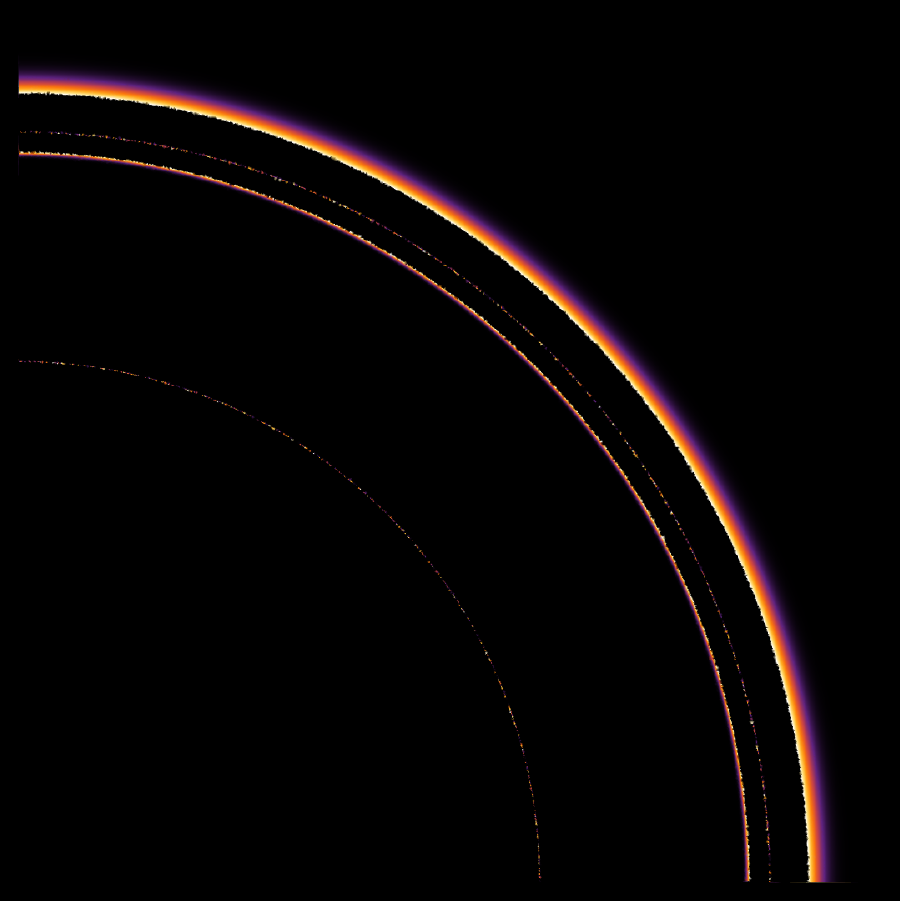}
\caption{Observed intensity and densityplot of emission model I. The top row corresponds to black hole and the bottom row corresponds to wormhole. The left collum are the observed intensities. The middle collum are the densityplots of observed intensities and the right collum are local densityplots.}\label{densityplot1}
\end{center}
\end{figure}
\begin{figure}[h!]
\begin{center}
\includegraphics[scale=0.45]{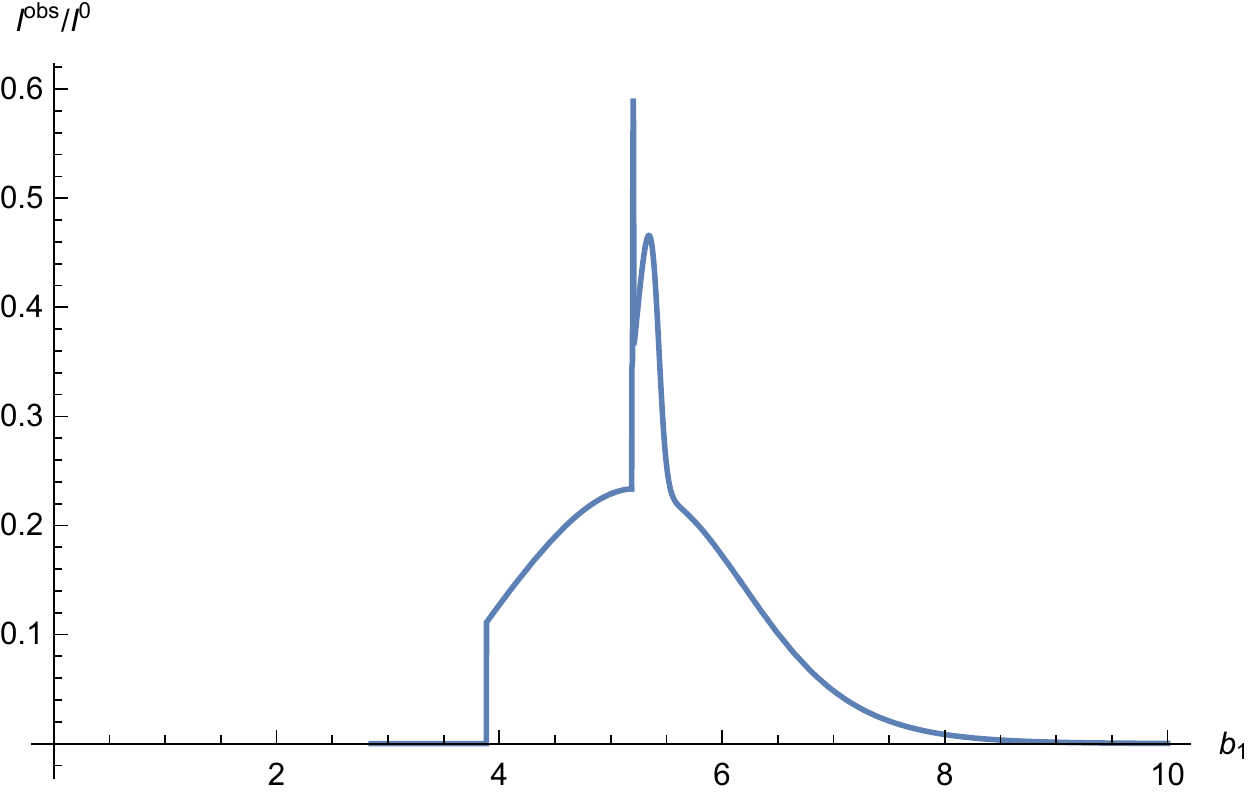}\qquad\includegraphics[scale=0.4]{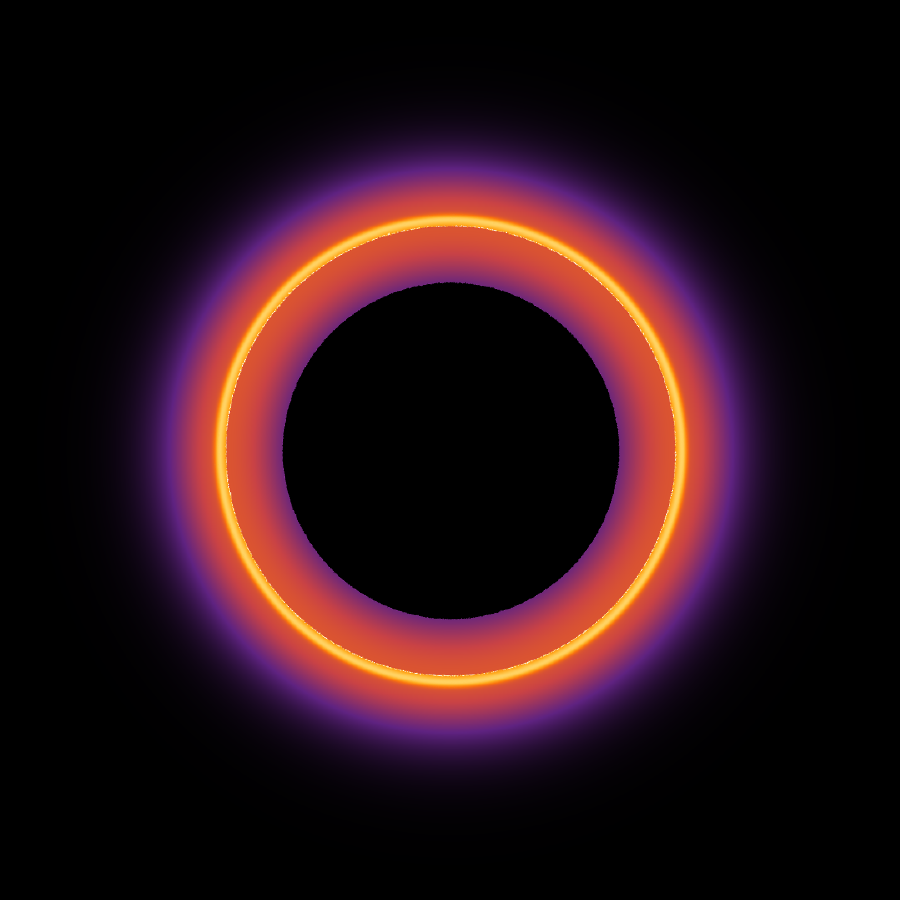}\qquad\includegraphics[scale=0.4]{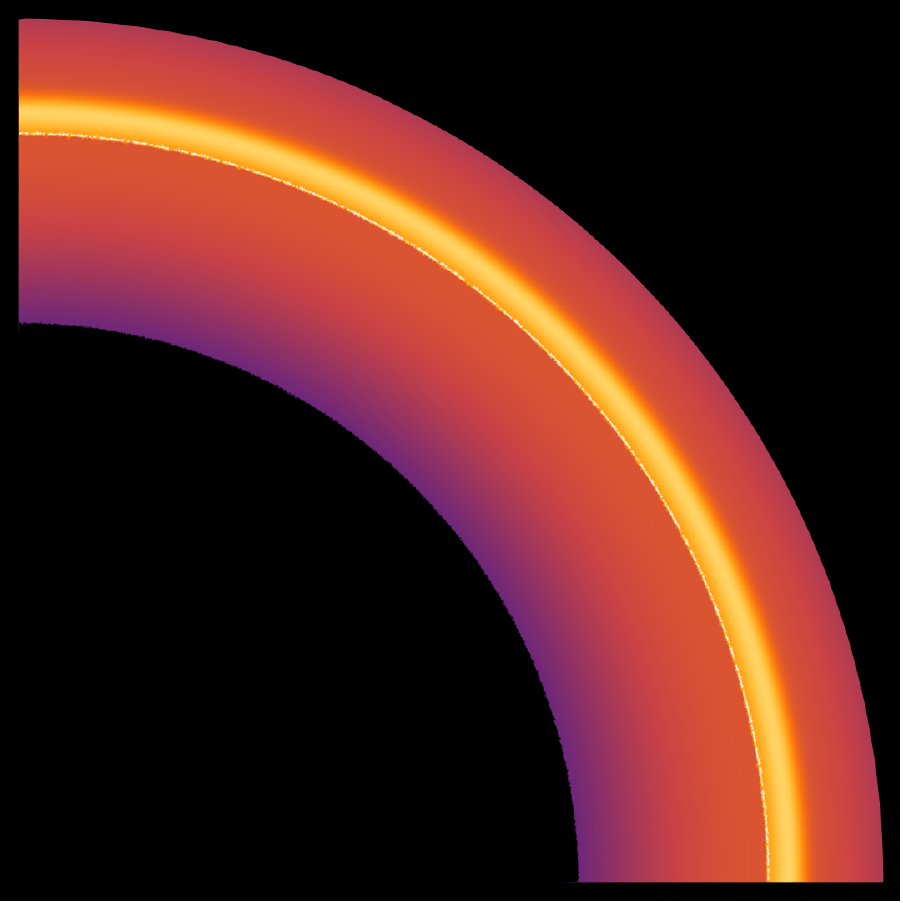}
\includegraphics[scale=0.45]{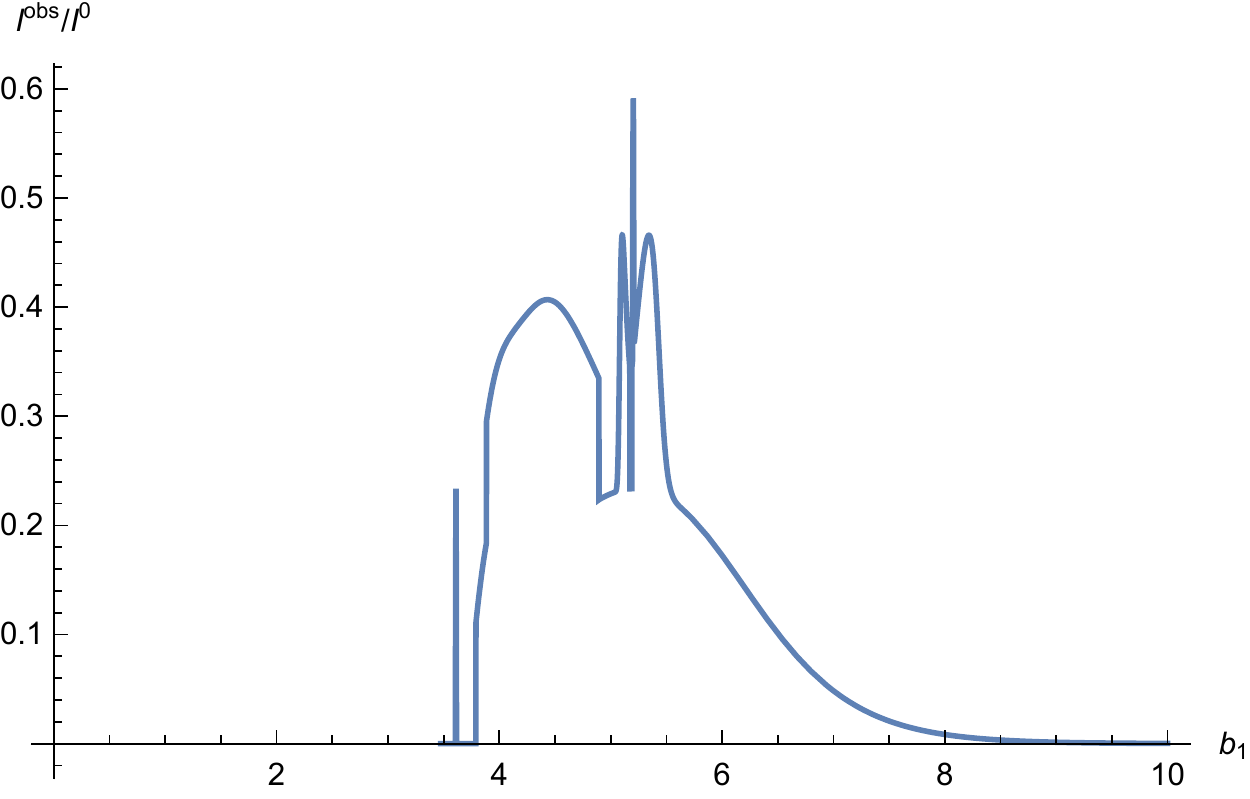}\qquad\includegraphics[scale=0.4]{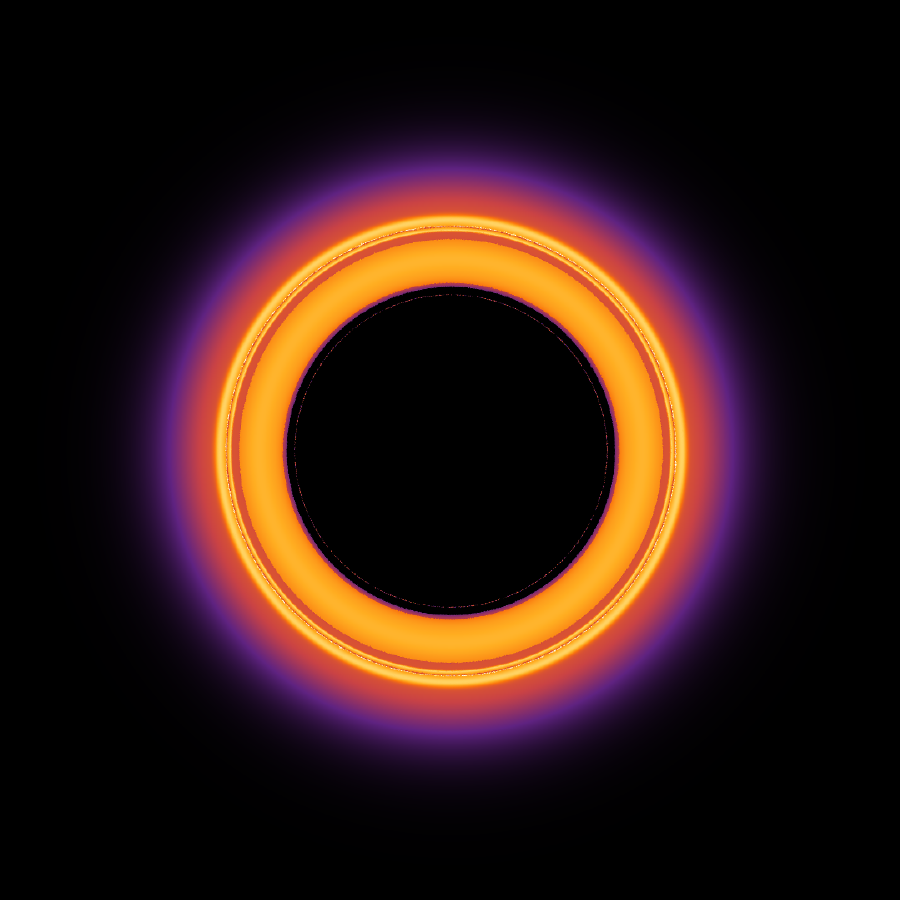}\qquad\includegraphics[scale=0.4]{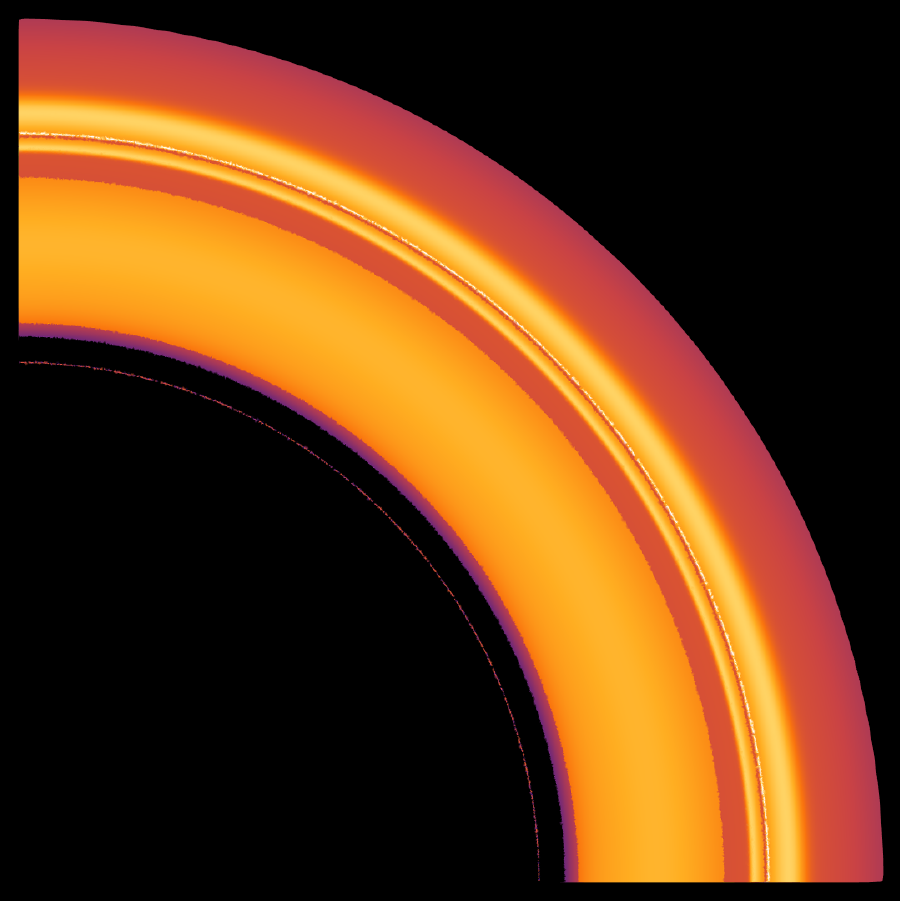}
\caption{Observed intensity and densityplot of emission model II. The top row corresponds to black hole and the bottom row corresponds to wormhole. The left collum are the observed intensities. The middle collum are the densityplots of observed intensities and the right collum are local densityplots.}\label{densityplot2}
\end{center}
\end{figure}

The observed intensity can be obtained according to \eqref{Iobs}. We give corresponding plots and densityplots of the observed intensities in Fig.\ref{densityplot1} for emission model I and Fig.\ref{densityplot2} for emission model II. We can see that in both models we have two additional photon rings resulted from corresponding two new third transfer functions for asymmetric shin shell wormhole. The new photon ring near critical curve $Zb_2^c$ is highly demagnified like the photon ring near critical curve $b_1^c$. Another new photon ring which is located at the inside of critical curve $b_1^c$ has a considerable size, but is smaller than the lensing ring which is located at outside of critical curve $b_1^c$. However in emission model I, the new second transfer function make no contribution to observed intensity, since this transfer function is out of the domain of emission model. While in emission model II, we have an additional sizable lensing band between critical curves $Zb_2^c$ and $b_1^c$ resulted from the new second transfer function in this emission model.

\section{Conclusion}
In this paper, we studied the trajectories of photons and their deflection angles in asymmetric thin-shell wormhole connecting two distinct Schwarzschild spacetimes by the throat. Typically, we placed observers in the spacetime $\mathcal{M}_1$ of which the mass parameter is smaller than that of the other side. Our interest is mainly about the photon rings in the observers' sky, we focused on the case that the throat of the ATW is inside the photon sphere of $\mathcal{M}_1$. After giving the formulas of deflection angles in each side, we constructed new orbit numbers counting the total deflection angles and showed the completed trajectories of photons which go through the throat twice, that is, ingoing photons of $\mathcal{M}_1$ pass through the throat and turn back after they reach the turning points in spacetime ${\cal M}_2$, then they go through the throat again. Then, considering optically and geometrically thin accretion disks around the wormhole, we gave the transfer functions and obtained the observed intensity and density plot in the sky of observers based on some emission models.

From our calculations, we found the ATW and corresponding Schwarzschild spacetimes differ markedly in the second and third transfer functions. In particular, the second transfer function is no longer a monotone function, a new segment appears before the monotone increasing part which also exists in corresponding Schwarzschild black hole spacetimes. As for the third transfer function,  more structures were found in ATW. In addition to a new monotone increasing part, the usual one splits into two branches. The underlying reasons for these new characteristics is that photons with certain impact parameters could turn back after they go through the throat while these photons in corresponding Schwarzschild black hole spacetimes would fall into the event horizon and never come back since the event horizon is a one-way membrane.

 As a result, two additional photon rings are found for ATW spacetime, one of which is highly demagnified near $Zb_2^c$ (the critical curve in opposite spacetime viewed by observers' side) like the usual photon ring near $b_1^c$ (the critical curve in observers' spacetime) and the other one located at inside of critical curve $b_1^c$ is much brighter and has a considerable size, even though it's smaller than the lensing ring which is located at outside of critical curve $b_1^c$. Besides, we also found an additional lensing band when the emission profile overlap the domain of the new second transfer function. Though the ATW is a constructed model, these additional photon rings or lensing rings (or bands) should be exclusive structures for ultra compact objects because of reflectivity \footnote{In fact, the relevant study of gravastar as another UCO example has been implemented in \cite{Wang:2020jek}.}. Our analysis provide an optically observational evidence to distinguish UCOs from black holes.

\section*{Acknowledgments}
J.P. is supported by the China Scholarship Council. M.G. is supported by China Postdoctoral Science Foundation Grant No. 2019M660278 and 2020T130020. X.H.F. is supported by NSFC (National Natural Science Foundation of China) Grant No. 11905157 and No. 11935009.

\end{document}